\documentclass[10pt,letterpaper]{article}
\usepackage{opex3}
\usepackage{graphicx}
\usepackage{pifont}
\usepackage{color}
\usepackage{url}
\usepackage{cite}
\usepackage{array}
\usepackage{hhline}
\usepackage[normalem]{ulem}
\newcommand{\circona}{\mbox{\textcircled{\tiny 1}}}
\newcommand{\circonb}{\mbox{\textcircled{\tiny 2}}}
\newcommand{\circonc}{\mbox{\textcircled{\tiny 3}}}
\newcommand{\circond}{\mbox{\textcircled{\tiny 4}}}

\begin{document}
\pagenumbering{arabic}\pagestyle{plain}

%\pagestyle{empty}
%\title{Control of optical spin Hall effect shift in  metasurface by weak measurement post-selection}
\title{Control of optical spin Hall shift in phase-discontinuity metasurface by weak value measurement post-selection }
\author{Y.U. Lee and J.W. Wu$^{*}$}

\address{Department of Physics and
Quantum Metamaterials Research Center\\
Ewha Womans University, Seoul 120-750, Korea}

\email{$^{*}$jwwu@ewha.ac.kr} %% email address is required

%

%\today
%\\

%\begin{abstract}
%\end{abstract}
%\ocis{(160.3918) Metamaterials;(130.5440) Polarization-selective devices.}

%%%%%%%%%%%%%%%%%%%%%%% References %%%%%%%%%%%%%%%%%%%%%%%%%

%%%%%%%%%%%%%%%%%%%%%%%%%%  body  %%%%%%%%%%%%%%%%%%%%%%%%%%
%
%Optical spin Hall effect (OSHE) originates from spin-orbit coupling present in an optical beam propagating through a curved trajectory.~\cite{Fedorov,imbert1972calculation}
%
%Weak measurement amplification has been employed to observe optical spin Hall  (OSH) shift in a refraction beam passing through an air-glass interface, the refractive index gradient $\vec{\nabla} n$ being normal to the interface.~\cite{hosten2008observation}
%
Optical spin Hall  (OSH) shift has been observed by weak measurement amplification in a refraction beam passing through air-glass interface, the refractive index gradient $\vec{\nabla} n$ being normal to the interface.~\cite{hosten2008observation}
%experiencing a gradient of refractive index.
Phase-discontinuity metasurface (PMS) possesses $\vec{\nabla} n$ tangential to the metasurface, and depending on the incidence angle either positive or negative refraction takes place satisfying the generalized Snell's law.~\cite{yu2011light}
Rapid phase-change over subwavelength distance at PMS leads to a large $\vec{\nabla} n$, enabling a direct observation of OSH shift.\cite{yin2013photonic}
%, a scheme holding in the weak interaction regime.\cite{yin2013photonic}
%
%
Here, we identify that the relative OSH shift between optical beams with spins $\pm 1$ depends on incidence and refraction angles at PMS,
%and construct a measurement of OSH effect with a variable phase retardance in the post-selection to demonstrate a control of transverse shift.
and demonstrate a control of OSH shift by constructing a weak value measurement with a variable phase retardance in the post-selection.
Capability of OSH shift control permits a tunable precision metrology applicable to nanoscale photonics such as angular momentum transfer and sensing.\\
%
%and beam refraction in LPGM follows the generalized Snell's law allowing  both positive and negative refractions. , and helicity-dependent transverse shifts are opposite for positive and negative refractions in LPGM.
%While OSHE in refraction at air/glass interface is detected by weak measurement amplification,
%transverse shit of refracted beam is lgives rise to OSHE
%incorporating the tangential momentum of $\vec{\nabla} n$.
% owing to a phase discontinuity $\vec{\nabla} \Phi$ present in the cross-polarized scattering light.
%, and  both positive and negative refractions takes place depending on incidence angle.
%Lorentz-like force on optical beam via Berry curvature in momentum space describes the transverse shift of OSHE.
%$\vec{\nabla} n$ being tangential to the metasurface,
%
%Here, we demonstrate a control of transverse shift, both sign and magnitude, by manipulating optical phase retardance in weak measurement post-selection.
%

%\section{Introduction}

%Optical spin Hall effect (OSHE) is an optics analogue of electron spin Hall effect, observed as a transverse spatial separation of two orthogonal circular polarized beams when a refractive index gradient is present in the interface of two media.
%

OSH effect is attributed to spin-orbit interaction of light in an optical beam propagating along a curved trajectory, originating from the transversality nature of electromagnetic field.~\cite{Fedorov,imbert1972calculation}
%
%In an inhomogeneous isotropic medium,
When a linearly polarized beam refracts at the interface of two optical media, optical beams with spins $\pm 1$ experience an OSH transverse spatial shift in opposite directions.\cite{onoda2004hall, bliokh2008geometrodynamics}
In a PMS composed of an array of V-shaped antennae, a large amount of refraction takes place in cross-polarized scattering light over subwavelength distance, leading to an OSH shift in extra-ordinary refraction beam in the order of a few hundreds nanometers at the near IR spectral range.\cite{yin2013photonic}

\begin{figure}[t]
\begin{center}
   \includegraphics[width=12cm]{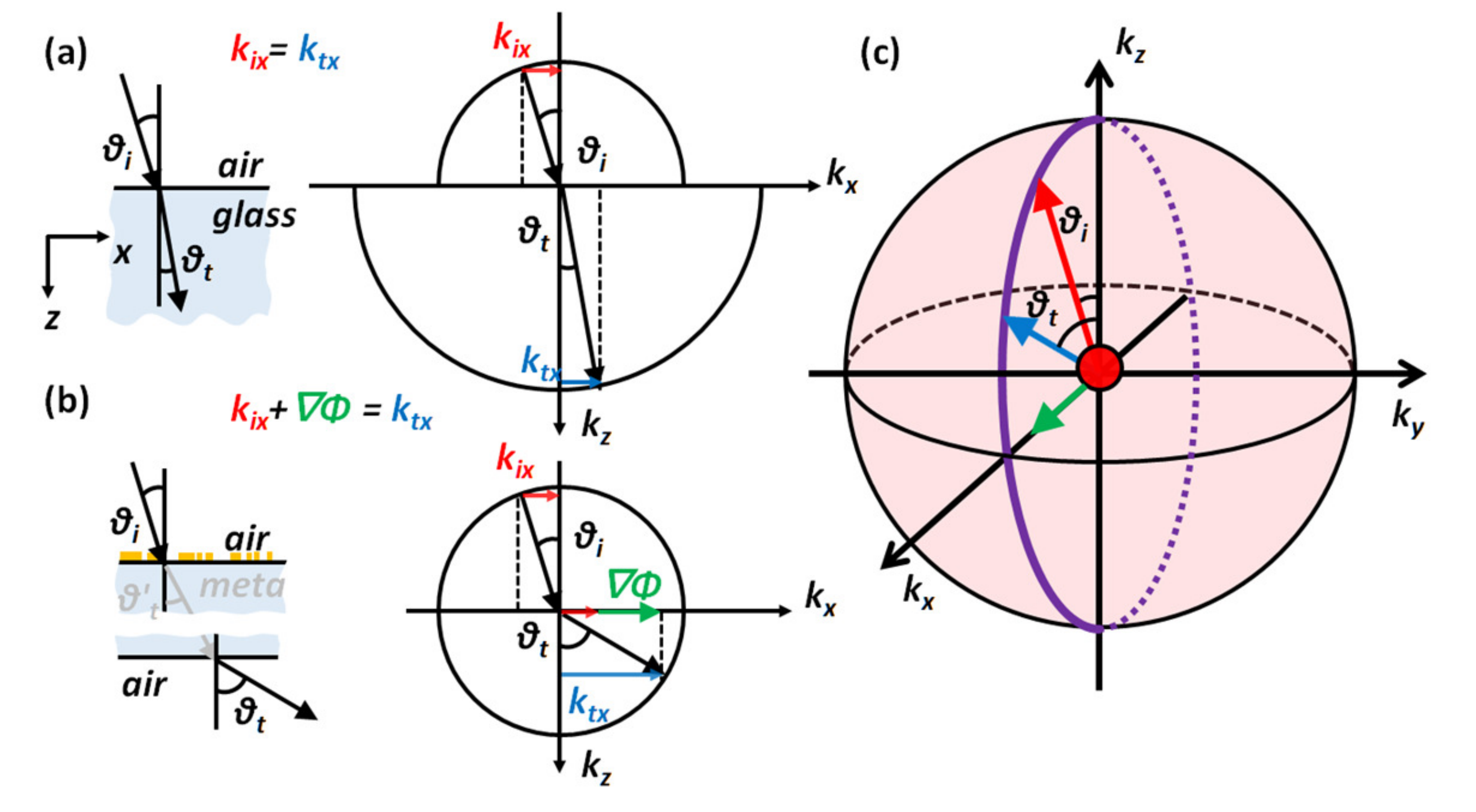}
    \caption
    {\label{Equifrequency}
    Schematics of optical beam refractions with refractive index gradient $\vec{\nabla} n$ in the incidence plane is shown.
   (a) At air-glass interface $\vec{\nabla} n$ is along $z$-axis normal to the surface, and radii of equifrequency surfaces are different in air and glass (b) At phase-discontinuity metasurface $\vec{\nabla} n$ is along $x$-axis tangential to the surface, and the net refraction after passing through the glass substrate is described by equifrequency surface in air with an additional momentum from $\vec{\nabla} \Phi$. (c) 3D plot of incidence (red) and refraction (blue) wavevectors and phase gradient $\vec{\nabla}\Phi$ (green) are shown. Monopole is located at the center of the equifrequency surface.
    %$\partial_{x}\Phi$ corresponds to the phase gradient in the metasurface.
    }
   \end{center}
\end{figure}

One distinct feature of PMS is that the refractive index gradient is tangential to the metasurface, differently from air-glass interface where the refractive index gradient is normal to the interface.
At the air-glass interface shown in Fig.~\ref{Equifrequency}(a), the radii of equifrequency surfaces are different in air and glass, and transverse shifts cancel out at top and bottom interfaces possessing opposite refractive index gradients.\cite{hosten2008observation}.
At PMS on a glass, on the other hand, the net transverse shift comes from the refractive index gradient of metasurface as shown in Fig.~\ref{Equifrequency}(b) with a single spherical equifrequency surface with the radius specified by the dispersion relation of light in air.\\

%See Fig.~\ref{Equifrequency}. To observe an OSH transverse shift at air/glass interface refraction, a structure of variable angle prism was adopted to avoid cancelation of transverse shifts at two interfaces possessing opposite refractive index gradient.\cite{hosten2008observation} In  PMS, on the other hand, the net OSH transverse shift comes from the refractive index gradient of metasurface.
%
%

%\section{Lorentz force in momentum space from the monopole Berry curvature}
The spin-orbit interaction of light is one example of interaction Hamiltonians coupling heavy/slow  and light/fast systems constituting a physical object. Coupling of heavy/slow and light/fast systems leads to effects of action and reaction between the two systems.~\cite{Berry-BookChapter,liberman1992spin,bliokh2008geometrodynamics}
Polarization-plane rotation of light along a coiled optical fiber results from the effect of a curved beam trajectory (slow) on optical spin (fast), which is a manifestation of the Berry phase in the light polarization.~\cite{PhysRevLett.57.933}
On the other hand, the effect of optical spin (fast) on a curved beam trajectory (slow) gives rise to a spin-dependent transverse shift of optical beam centroid, described by the Lorentz force in momentum space, $\lambda {\vec{\nabla}n} \times{{\vec{p}}\over p^3}$, where ${{\vec{p}}\over p^3}$ is the monopole Berry curvature associated with optical beam of spin $\lambda$.\cite{onoda2004hall,onoda2006geometrical,bliokh2008geometrodynamics,bliokh2009geometrodynamics}

In the equifrequency surface of PMS shown in Fig.~\ref{Equifrequency}(c) the refractive index gradient is along $x$-axis, and the monopole Berry curvatures are radial vectors with directions determined by incidence and refraction angles $\theta_i$ and $\theta_t$.
%
%A relation between the transverse shift and a phase gradient residing in the metasurface.
Transverse shift $\delta {{y}}$  upon refraction at PMS is related to the phase gradient $\vec{\nabla}\Phi= {{2\pi}}\vec{\nabla} n$ and Berry connections of incidence and refraction beams, yielding the expression of transverse shift:~\cite{onoda2006geometrical,bliokh2005topological}
\begin{eqnarray}
\delta {{y}}
%&=&\langle{{z}^{i}}\left| {{\Lambda }_{{{k}^{i}}}} \right|{{z}^{i}}\rangle-\langle{{z}^{t}}\left| {{\Lambda }_{{{k}^{t}}}} \right|{{z}^{t}}\rangle \nonumber\\
%&=&\mp \left( \frac{{{p}_{xt}}{{p}_{zt}}}{{{p}_{t}}(p_{xt}^{2}+p_{yt}^{2})}-\frac{{{p}_{xi}}{{p}_{zi}}}{{{p}_{i}}(p_{xt}^{2}+p_{yt}^{2})} \right) \nonumber \\
&=&-\lambda \frac{\cos {{\theta }_{t}}-\cos {{\theta }_{i}}}{\vert{\vec{\nabla} \Phi }\vert }~.
{\label{meta}}
%\\{{\Lambda }_{k}}=-\frac{\cos \theta }{k\sin \theta }{{\sigma }_{3}}{{\textbf{e}}_{\phi }}
\end{eqnarray}
See SI for derivation of Eq.~(\ref{meta}).
%where $ |{{z}^{t,i}}\rangle$ and ${{\Lambda }_{{{k}^{t,i}}}}$ correspond to polarization states and Berry connections for refracted and incident beams.~\cite{bliokh2005topological}
%
%
From the two facts that both positive and negative refractions can take place at PMS and that $\vec{\nabla} n$ is tangential to the PMS surface,
the sign and magnitude of transverse shift $\delta {{y}}$ depend on incidence and refraction angles $\theta_i$ and $\theta_t$ as well as $\vert{\vec{\nabla} \Phi }\vert$, as can be read-off  from Fig.~\ref{Equifrequency}(c) and Eq.~(\ref{meta}).

\begin{figure}[t]
\begin{center}
   \includegraphics[width=14cm]{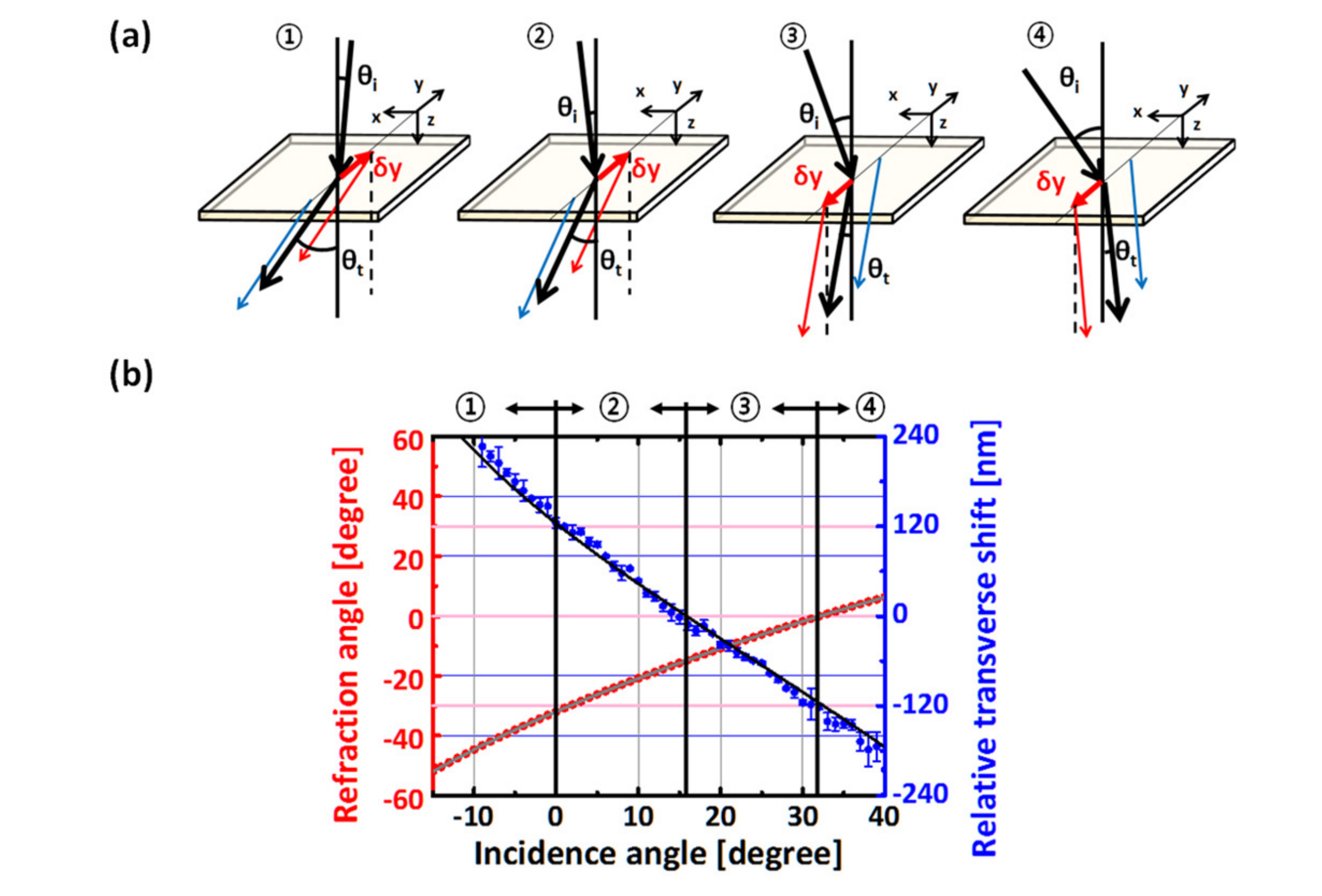}
     \caption
      {\label{PRLM}
   (a) Schematics of the beam refraction from air to  \circona ~a positive-refraction-low-index medium, \circonb ~a negative-refraction-low-index medium, \circonc ~a negative-refraction-high-index medium, and \circond ~a positive-refraction-high-index medium are shown.
   (b) Theoretical calculation (solid curve) and experimental measurement (solid circle) of refraction angle (red) $\theta_t$ and relative transverse shift  (blue) are plotted as a function of incidence angle $\theta_i$.
   }
   \end{center}
\end{figure}

Figure~\ref{PRLM}(a) shows examples how the relative transverse shift of optical beams with spins $\pm 1$ changes sign in detail.
For $\lambda = +1$ corresponding to the red arrows in Fig.~\ref{PRLM}(a), when $\theta_i < \theta_t$ a positive transverse shift ($\delta {{y}}>0$) takes place in both positive (\circona~) and negative (\circonb~) refractions, and when $\theta_i > \theta_t$ a negative transverse shift ($\delta {{y}}<0$) takes place in both negative (\circonc~) and positive (\circond~) refractions.
In Fig.~\ref{PRLM}(b) are plotted theoretical calculation (solid curves) and experimental measurement (solid circles) of refraction angle $\theta_t$ and transverse shift $\delta y$  as a function of incidence angle $\theta_i$.\\

%\section{Weak value of OSH shift post-selected with a phase retardance}
%
Weak measurement amplification enabled the observation of OSH shift in air-glass interface.~\cite{hosten2008observation}
By preparing a polarizer as pre-selection, the weak value is measured by a strong measurement with a nearly cross-polarized analyzer as post-selection.~\cite{ritchie1991realization}
In PMS, on the other hand, OSH shift was directly detected without resorting to a weak measurement amplification.
%According to the analogy between optical beam shifts and quantum weak measurements, OSH shift is a classical analogue of a quantum measurement of the polarization state of a paraxial beam by its transverse amplitude distribution.~\cite{dennis2012analogy}
%
%However, in order to control the transverse shift in PMS by an optical means, a weak value measurement can be constructed with a variable phase retardance in the post-selection.
%
When it is attempted to control the transverse shift in PMS by an optical means, however, a weak value measurement can be utilized with a variable phase retardance in the post-selection.

\begin{figure}[t]
\begin{center}
   \includegraphics[width=12cm]{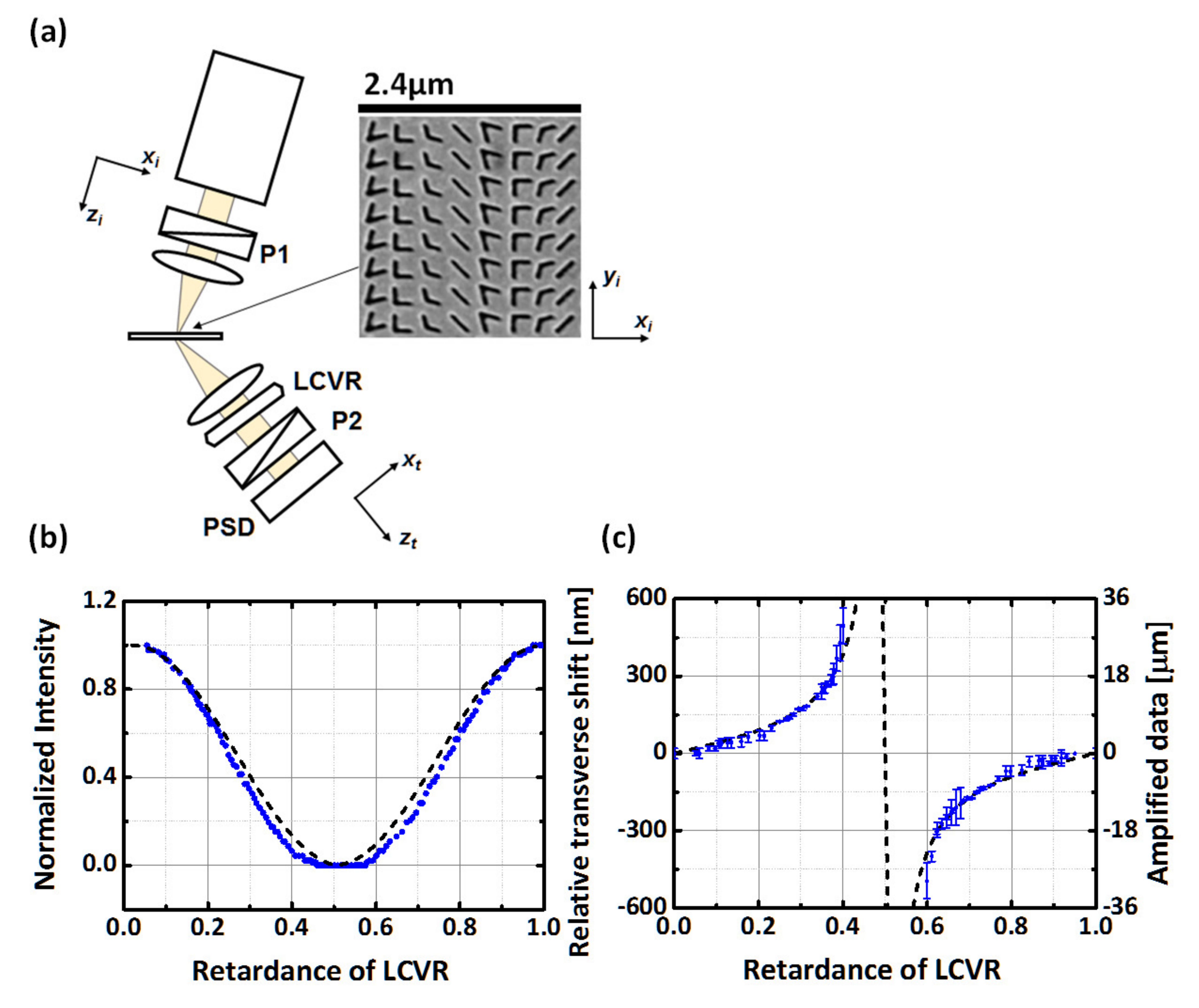}
     \caption
      {\label{setup2}
  (a)  Schematics of weak measurement with a variable retardance is shown with $P_1$=(1,0)$^T$ and $P_2$=(0,1) along with SEM image of Babinet complementary phase-gradient metasurface.~\cite{Lee:14}  LCVR is liquid-crystal variable retarder and PSD is a quadrant position sensitive detector. See Methods for the detailed description of sample and measurement. (b) Light intensity transmitted through a cross polarizer/analyser setup is plotted as a function of retardance $\Gamma$ of LCVR. (c)  Weak value of OSHE shift post-selected with phase retardance is plotted as a function of retardance $\Gamma$ of LCVR with the corresponding transverse shift. Blue solid circles are data point and dashed curves are from theoretical calculation. See SI for images of spin-dependent OSH shifts.
}
   \end{center}
\end{figure}

OSH shift is one example of classical analogues of a quantum measurement of the polarization state of a paraxial beam by its transverse amplitude distribution.~\cite{dennis2012analogy}
%A linearly polarized paraxial beam plays the role of the pre-selected quantum product state, and the spatial amplitude distribution and the homogeneous polarization correspond to a measuring device (pointer) and an object to be measured, respectively.
%
By introducing a variable optical phase retardance in the post-selection, we can tune the post-selected state $\vert\psi_f >$
across the whole range of retardance, [$0$, $\pi/2$], to control OSH shift, which is made possible in PMS since OSH shift is large enough to be observed in  the optical far field.
We place a phase retarder with variable retardance $\Gamma$ (modulus of $\pi$) inside the cross polarizer/analyzer ($P_1 / P_2$) setup in order to control OSH shift in the weak measurement as shown in Fig.~\ref{setup2}, where the post-selection state is
$\vert \psi_f(\Gamma)> = \big(\cos(\pi\Gamma), -i\sin(\pi\Gamma) \big)$.
%with Jones vectors of polarizer $P_1$=(1,0)$^T$ and analyzer $P_2$=(0,1).
%

%Refraction satisfying generalized Snell's law comes from a cross-polarized scattering light from an array of V-shaped antennas in the phase-grade metasurface. Hence polarization states of refraction and incidence beams are opposite.
%
%With a variable phase retardance $\Gamma$ is positioned inside the parallel-polarized setup,

%
When OSH transverse shift is measured at the propagation distance $z$ of a Gaussian beam with Rayleigh range of $z_0$, the observable $metaOSH$ is expressed in terms of the Pauli matrix $\hat{\sigma}_2$  in the linear polarization bases:~\cite{jayaswal2014observation,bliokh2005topological,toppel2013goos}
\begin{equation}
metaOSH=-
  \hat{\sigma}_2
\cdot \delta y \cdot \frac{z}{z_{0}}~.
\end{equation}
The weak value of transverse shift, post-selected at a retardance $\Gamma$, is readily obtained.
 \begin{equation}
\delta y_{w}(\Gamma)
=\frac{\left\langle  \psi(\Gamma)  \right|metaOSH\left| P_1 \right\rangle }
{\left\langle  \psi(\Gamma)  \right|\left. P_1 \right\rangle }
=  {z\over{z_0}}\cdot {{\delta y}\cdot {\tan\big(\pi\Gamma\big)}}
\end{equation}
%
%where $z = f = 60 z_0$ in our experiment.
Note that the phase retardance $\Gamma= {1\over 2}\pm \epsilon$ ($0<\epsilon \ll 1$) is the range where a weak measurement amplification is achieved.
In Fig.~\ref{setup2}(b) is plotted the transmitted light intensity through the cross polarizer/analyser setup of  Fig.~\ref{setup2}(a) as a function of retardance $\Gamma$.
Figure~\ref{setup2}(c) shows the weak value, $\delta y_{w}(\Gamma)$, of an optical beam normally incident on PMS
%of OSHE shift post-selected at a retardance $\Gamma$
as a function of retardance $\Gamma$ along with the corresponding transverse shift $\delta y$.
At $\Gamma=1/4$ the weak value $\delta y_{w}(\Gamma=1/4)= 7.44\mu m$, which corresponds to the transverse shift $\delta y = 124nm$ in the absence of  cross-polarized polarizer/analyzer.
It is important to note that the phase-retardance dependent weak value is measured in the far field.~\cite{gorodetski2012weak,dressel2014colloquium,kofman2012nonperturbative}\\
%

%\section{Control of OSHE transverse shift by weak measurement post-selection}
\begin{figure}[t]
\begin{center}
   \includegraphics[width=14cm]{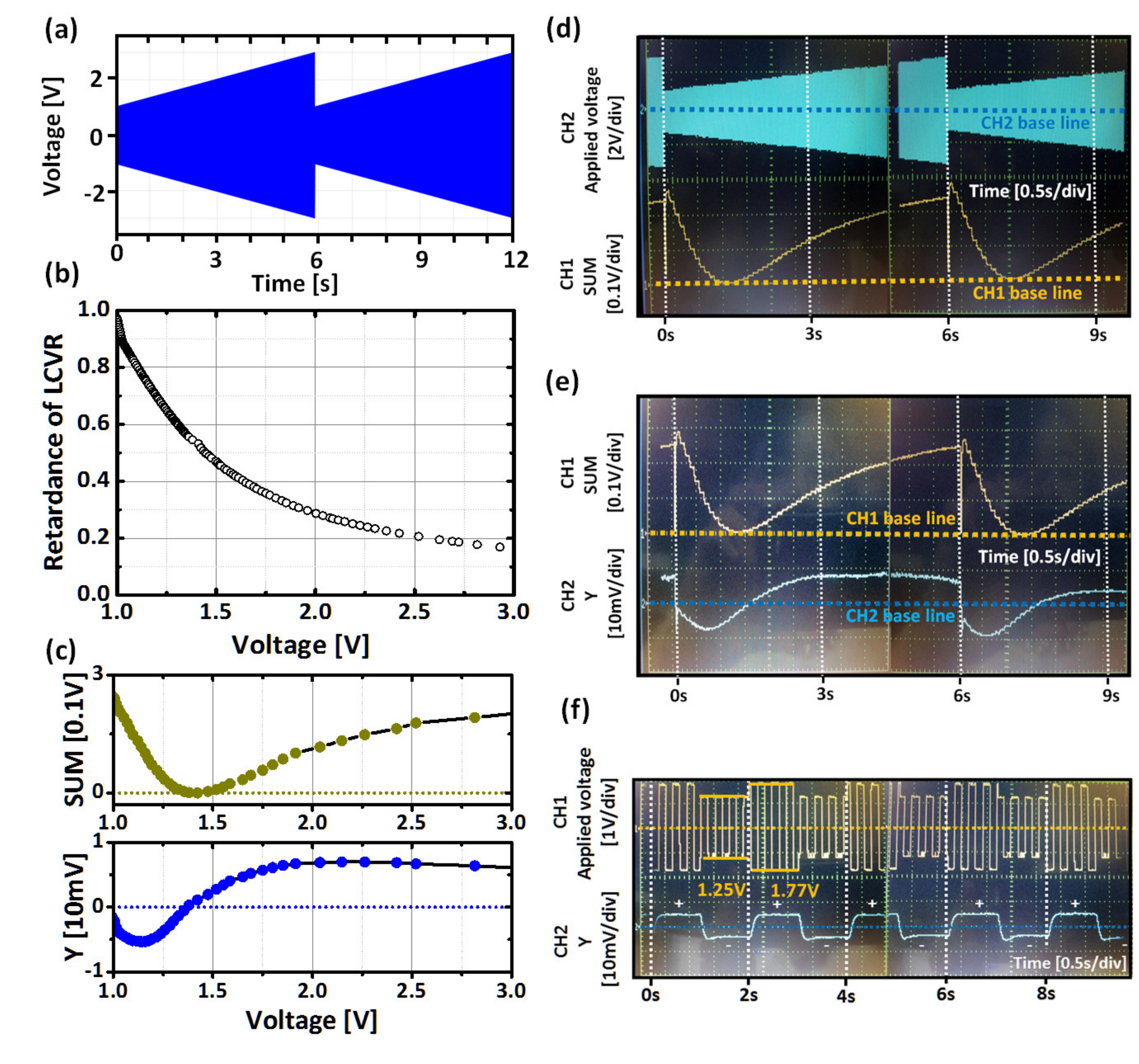}
     \caption
      {\label{osc2}
      (a) Saw-tooth waveform of LCVR driving voltage is plotted. (b) Retardance of LCVR and (c) SUM and Y of PSD are plotted as a function of LCVR driving voltage. Oscilloscope traces of (d) saw-tooth waveform and SUM of PSD and (e) SUM and Y of PSD are shown. (f) Switching between positive and negative Y is demonstrated as driving voltage is varied.
     }
   \end{center}
\end{figure}

%, not limited to the weak interaction regime of weak measurement.~\cite{gorodetski2012weak,dressel2014colloquium,kofman2012nonperturbative}
%

%\section{Control of OSHE transverse shift by weak measurement post-selection}

Since the weak value is  post-selected at phase retardance $\Gamma$, an electric manipulation of phase retardance in LCVR allows a control of the weak value $\delta y_{w}$. %measured in the strong interaction regime.

Figure~\ref{osc2} shows plots and oscilloscope traces of OSH shift control. A saw-tooth waveform of LCVR driving voltage is adopted with $1.0V$  and $3.0V$ as the initial and final voltages, which covers the phase retardance from 0 to 1 (modulus of $\pi$). From a position-sensitive detector (PSD), SUM =$q_{1}+ q_{2}+q_{3}+q_{4}$ and
Y=$(q_{1}+ q_{2})-(q_{3}+q_{4})$ are monitored, where Y is associated with OSH shift.

Comparison of Fig.~\ref{osc2} (b) and (c) shows that a sign reversal in Y takes place across the phase retardance $\Gamma = {1\over 2}$, at the vicinity of which a weak measurement amplification is achieved shown in Fig.~\ref{setup2}(b) and (c).
This leads to a switching behavior of post-selected OSH transverse shift when the phase retardance is varied crossing $\Gamma = {1\over 2}$.
In Fig.~\ref{osc2}(f) is demonstrated a switching between positive and negative Y as the driving voltage is alternated between $1.25V$ ($\Gamma = 0.65$) and $1.77V$ ($\Gamma = 0.35$). Switching operation of post-selected OSH shift has a strong implication in nanoscale photonics applications such as angular momentum transfer and sensing.\\

\textbf{Methods}\\

\textbf{Sample fabrication.} Phase-discontinuity metasurface is composed of V-shape antenna pattern.~\cite{yu2011light}
A linear array of eight V-shape apertures is repeated along $x$-axis with the lattice constant $\Gamma$¥Ã of 2400nm. Focused ion beam milling is utilized to fabricate Babinet complementary V-shaped antennas on e-beam evaporated 30nm-thick Au film on top of fused silica substrate with adhesion layer of 3nm thick titanium.~\cite{Lee:14}\\

\textbf{Experimental set-up}. We adopted $10 mW$ $\lambda=1310nm$ pigtail style self-contained thermally stabilized laser diode as the light source (OZ optics-OZ-2000) with the output fiber diameter 50 $\mu$m. The beam passes through a Glan/Thomson polarizer $P1$ (Thorlabs-GL10-C) to be linearly polarized. Then it is focused onto the metasurface with a microscope objective lens, $f=95mm$, to a $1/e^2$ intensity spot size $w_0$ = $50\mu m$. The extraordinary refraction beam is collected with a microscope objective lens, $f=95mm$, and a liquid crystal variable retarder (Thorlabs-LCC1113-C) and a second polarizer $P2$ are adopted to resolve the polarization state with an InGaAs-based NIR camera (Ophir-XC-130), and InGaAs-based quadrant position sensitive detector (Newport-2903) with a 3-mm diameter active region is employed for imaging and detection. In our experimental set-up, the propagation distance is $z = f = 60~ z_0$.
%The relative transverse shift is measured by varying the incidence angle of beam. To improve S/N ratio the detector is shielded from stray room lights.
%The polarization of the incidence is linear and can be adjusted in either the $x$ or $y$ directions with a half wave plate.
Position sensitive detector is connected to the oscilloscope, and the position X, Y, and SUM data are monitored. %We used position transfer function to measure the beam position. %The transfer function is related to the
The relative transverse shift [nm] $= %(Y_{\sigma^{+}}/SUM_{\sigma^{+}}-Y_{\sigma^{-}}/SUM_{\sigma^{-}})\times 900~[nm]$.
(Y_{\sigma^{+}}/SUM_{\sigma^{+}}-Y_{\sigma^{-}}/SUM_{\sigma^{-}})\times 1.08 \times beam~ radius ~[nm]$, which is obtained from the light intensity measurement by a photo-reciver placed on two dimensional translation stage.\\

\textbf{Acknowledgement}
This work is supported by Quantum Metamaterial Research Center program and Global
Frontier Program of Center for Advanced Meta-Materials (Ministry of Science, ICT and
Future Planning, National Research Foundation, Republic of Korea).\\

\newpage

\textbf{\centerline{SUPPLEMENTARY INFORMATION}}\\

\textbf{Control of optical spin Hall shift in phase-discontinuity metasurface by weak value measurement post-selection}

%\author{Y.U. Lee and J.W. Wu$^{*}$}
\author{Y.U. Lee and J.W. Wu}

%\address{Department of Physics and
%
%Quantum Metamaterials Research Center\\
%
%Ewha Womans University, Seoul 120-750, Korea}

%\email{$^{*}$jwwu@ewha.ac.kr} %% email address is required
\vspace{1cm}

Supplementary information consists of S1. Derivation of OSH shift in terms of Berry connections in PMS and S2. Images of spin-dependent OSH shifts.\\

%
%\bibliographystyle{osajnl}
%\bibliography{OSH-06Jan2015-ref}

\newpage
\begin{thebibliography}{10}
\newcommand{\enquote}[1]{``#1''}

\bibitem{hosten2008observation}
O.~Hosten and P.~Kwiat, \enquote{Observation of the spin Hall effect of light
  via weak measurements,} Science \textbf{319}, 787--790 (2008).

\bibitem{yu2011light}
N.~Yu, P.~Genevet, M.~A. Kats, F.~Aieta, J.-P. Tetienne, F.~Capasso, and
  Z.~Gaburro, \enquote{Light propagation with phase discontinuities:
  generalized laws of reflection and refraction,} Science \textbf{334},
  333--337 (2011).

\bibitem{yin2013photonic}
X.~Yin, Z.~Ye, J.~Rho, Y.~Wang, and X.~Zhang, \enquote{Photonic spin Hall
  effect at metasurfaces,} Science \textbf{339}, 1405--1407 (2013).

\bibitem{Fedorov}
F.~I. Fedorov, \enquote{Theory of total reflection,} Dokl. Akad. Nauk SSSR
  \textbf{105}, 465?--468 (1955).

\bibitem{imbert1972calculation}
C.~Imbert, \enquote{Calculation and experimental proof of the transverse shift
  induced by total internal reflection of a circularly polarized light beam,}
  Physical Review D \textbf{5}, 787 (1972).

\bibitem{onoda2004hall}
M.~Onoda, S.~Murakami, and N.~Nagaosa, \enquote{Hall effect of light,} Physical
  Review Letters \textbf{93}, 083901 (2004).

\bibitem{bliokh2008geometrodynamics}
K.~Y. Bliokh, A.~Niv, V.~Kleiner, and E.~Hasman, \enquote{Geometrodynamics of
  spinning light,} Nature Photonics \textbf{2}, 748--753 (2008).

\bibitem{Berry-BookChapter}
M.~Berry, \enquote{The quantum phase, five years after,} in \enquote{Geometric
  phases in physics,} , D.~L. Andrews and M.~Babiker, eds. (World Scientific
  Singapore, 1989).

\bibitem{liberman1992spin}
V.~Liberman and B.~Y. Zel¡¯dovich, \enquote{Spin-orbit interaction of a photon
  in an inhomogeneous medium,} Physical Review A \textbf{46}, 5199 (1992).

\bibitem{PhysRevLett.57.933}
R.~Y. Chiao and Y.-S. Wu, \enquote{Manifestations of Berry's topological phase
  for the photon,} Physical Review Letters \textbf{57}, 933--936 (1986).

\bibitem{onoda2006geometrical}
M.~Onoda, S.~Murakami, and N.~Nagaosa, \enquote{Geometrical aspects in optical
  wave-packet dynamics,} Physical Review E \textbf{74}, 066610 (2006).

\bibitem{bliokh2009geometrodynamics}
K.~Y. Bliokh, \enquote{Geometrodynamics of polarized light: Berry phase and
  spin Hall effect in a gradient-index medium,} Journal of Optics A: Pure and
  Applied Optics \textbf{11}, 094009 (2009).

\bibitem{bliokh2005topological}
K.~Y. Bliokh and V.~Freilikher, \enquote{Topological spin transport of photons:
  Magnetic monopole gauge field in Maxwell's equations and polarization
  splitting of rays in periodically inhomogeneous media,} Physical Review B
  \textbf{72}, 035108 (2005).

\bibitem{ritchie1991realization}
N.~Ritchie, J.~Story, and R.~G. Hulet, \enquote{Realization of a measurement of
  a ¡®¡®weak value¡¯¡¯,} Physical Review Letters \textbf{66}, 1107 (1991).

\bibitem{dennis2012analogy}
M.~R. Dennis and J.~B. G{\"o}tte, \enquote{The analogy between optical beam
  shifts and quantum weak measurements,} New Journal of Physics \textbf{14},
  073013 (2012).

\bibitem{jayaswal2014observation}
G.~Jayaswal, G.~Mistura, and M.~Merano, \enquote{Observation of the
  Imbert--Fedorov effect via weak value amplification,} Optics Letters
  \textbf{39}, 2266--2269 (2014).

\bibitem{toppel2013goos}
F.~T{\"o}ppel, M.~Ornigotti, and A.~Aiello, \enquote{Goos--h{\"a}nchen and
  Imbert--Fedorov shifts from a quantum-mechanical perspective,} New Journal of
  Physics \textbf{15}, 113059 (2013).

\bibitem{Lee:14}
Y.~U. Lee, J.~Kim, J.~H. Woo, L.~H. Bang, E.~Y. Choi, E.~S. Kim, and J.~W. Wu,
  \enquote{Electro-optic switching in phase-discontinuity complementary
  metasurface twisted nematic cell,} Optics Express \textbf{22}, 20816--20827
  (2014).

\bibitem{gorodetski2012weak}
Y.~Gorodetski, K.~Bliokh, B.~Stein, C.~Genet, N.~Shitrit, V.~Kleiner,
  E.~Hasman, and T.~Ebbesen, \enquote{Weak measurements of light chirality with
  a plasmonic slit,} Physical Review Letters \textbf{109}, 013901 (2012).

\bibitem{dressel2014colloquium}
J.~Dressel, M.~Malik, F.~M. Miatto, A.~N. Jordan, and R.~W. Boyd,
  \enquote{Colloquium: Understanding quantum weak values: Basics and
  applications,} Reviews of Modern Physics \textbf{86}, 307 (2014).

\bibitem{kofman2012nonperturbative}
A.~G. Kofman, S.~Ashhab, and F.~Nori, \enquote{Nonperturbative theory of weak
  pre-and post-selected measurements,} Physics Reports \textbf{520}, 43--133
  (2012).


\bibitem{onoda2006geometrical}
M.~Onoda, S.~Murakami, and N.~Nagaosa, \enquote{Geometrical aspects in optical
  wave-packet dynamics,} Physical Review E \textbf{74}, 066610 (2006).

\bibitem{bliokh2005topological}
K.~Y. Bliokh and V.~Freilikher, \enquote{Topological spin transport of photons:
  Magnetic monopole gauge field in Maxwell's equations and polarization
  splitting of rays in periodically inhomogeneous media,} Physical Review B
  \textbf{72}, 035108 (2005).
\end{thebibliography}

\textbf{S1. Derivation of OSH shift in terms of Berry connections in PMS}\\

The transverse shift upon refraction has been related to the Berry connection.~\cite{onoda2006geometrical,bliokh2005topological}
\begin{eqnarray}
\delta {{y}}=\langle{{z}^{i}}\left| {{\Lambda }_{{{k}^{i}}}} \right|{{z}^{i}}\rangle-\langle{{z}^{t}}\left| {{\Lambda }_{{{k}^{t}}}} \right|{{z}^{t}}\rangle
%\\{{\Lambda }_{k}}=-\frac{\cos \theta }{k\sin \theta }{{\sigma }_{3}}{{\textbf{e}}_{\phi }}
\end{eqnarray}
where $ |{{z}^{t,i}}\rangle$ and ${{\Lambda }_{{{k}^{t,i}}}}$ stand for polarization state and Berry connection of refraction and incidence beams.
Bliokh \textit{et al}. expressed the Berry connection ${{{\hat{A}}}}^{(\lambda)}$ in terms of rectangular components of the linear momentum to obtain OSH shift $\delta {{y}}$ of optical beam with spin $\lambda$.~\cite{bliokh2005topological}
For PMS surface, we have
\begin{eqnarray}
%\left( {{{\Lambda}}_{yt}}^{\pm }-{{{\Lambda}}_{yi}}^{\pm } \right)
%&=&\mp \left( \frac{{{p}_{zt}}}{{{p}_{t}}{{p}_{xt}}}-\frac{{{p}_{zi}}}{{{p}_{i}}{{p}_{xi}}} \right)  \nonumber \\
\delta {{y}}&=& {{{\hat{A}}}_{iy}}^{(\lambda)}-{{{\hat{A}}}_{ty}}^{(\lambda)}\nonumber \\
&=&\lambda \left( \frac{{{p}_{tx}}{{p}_{tz}}}{{{p}_{t}}(p_{tx}^{2}+p_{ty}^{2})}-\frac{{{p}_{ix}}{{p}_{iz}}}{{{p}_{i}}(p_{ix}^{2}+p_{iy}^{2})} \right) \nonumber \\
&=&\lambda \left( \frac{{{p}_{tz}}}{{{p}_{t}}{{p}_{tx}}}-\frac{{{p}_{iz}}}{{{p}_{i}}{{p}_{ix}}} \right)\nonumber \\
%&=&\mp \left( \frac{\cos {{\theta }_{t}}}{\nabla \Phi }-\frac{\cos {{\theta }_{i}}}{\nabla \Phi } \right) \nonumber \\
&=&-\lambda \frac{\cos {{\theta }_{t}}-\cos {{\theta }_{i}}}{\vert\nabla \Phi\vert }
\end{eqnarray}
where ${p}_{tx}=-\hbar\vert\nabla\Phi\vert$, ${p}_{tz}=\hbar k_{t}\cos\theta_t$, ${p}_{i}={p}_{t}=\hbar k_{i}=\hbar k_{t}$, ${p}_{ix}=-\hbar\vert\nabla\Phi\vert$, ${p}_{iz}=k_{i}\cos\theta_i$, $\nabla\Phi=-\vert\nabla\Phi\vert\hat{x}$, and it is noted that ${p}_{iy}={p}_{ty}=0$.\\

\textbf{S2. Images of spin-dependent OSH shifts } \\

\begin{figure}[t]
\begin{center}
\includegraphics[width=13cm]{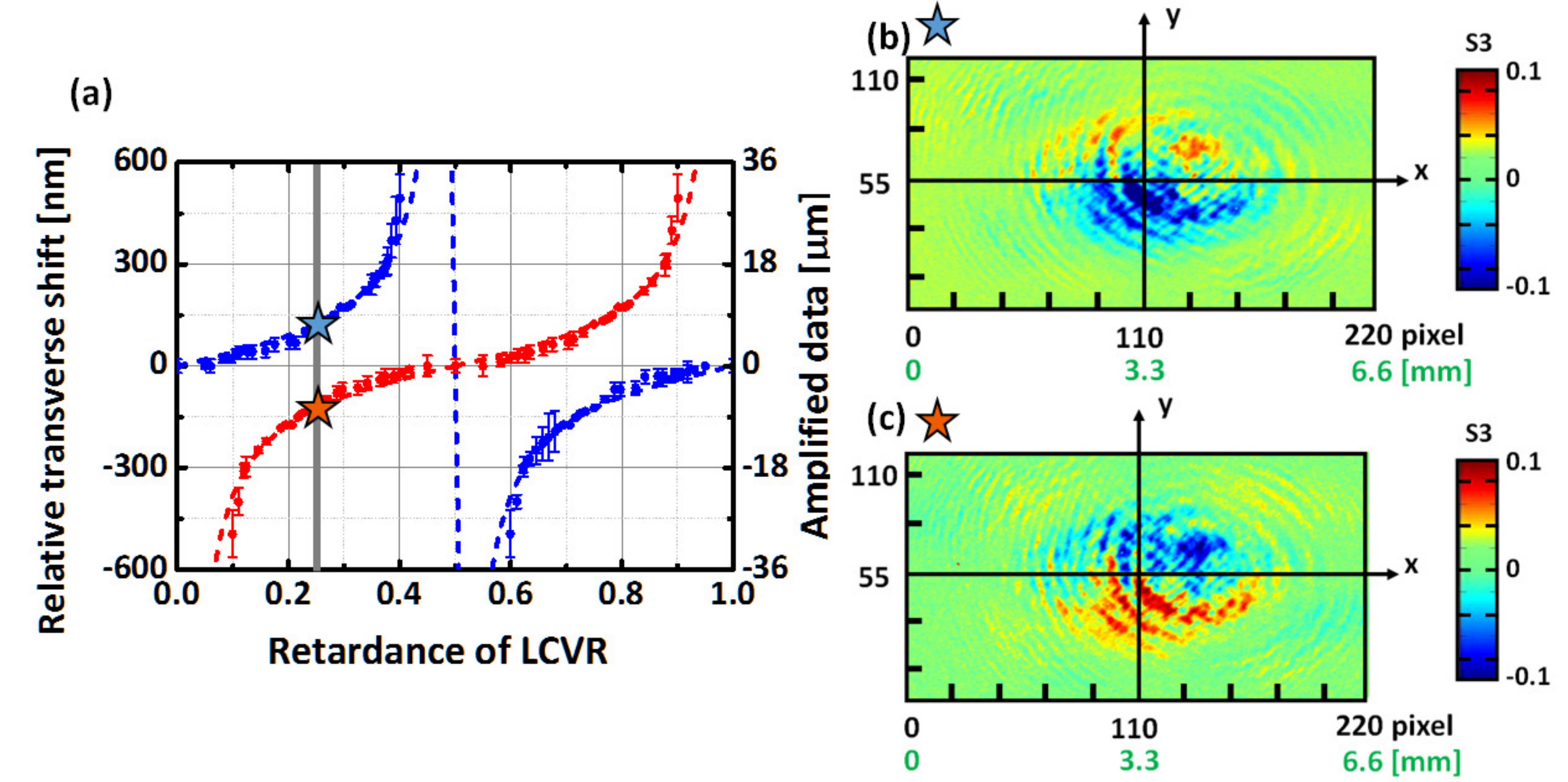}
     \caption
      {\label{S1}
   (a) Relative transverse shifts are measured as a function of retardance $\Gamma$ in cross-polarized polarizer/analyser setup (blue solid circles) and in parallel-polarized polarizer/analyser setup (red solid circles). Images of spin-dependent OSH shifts at $\Gamma=1/4$ (vertical gray straight line in (a)) are obtained by processing each pixel signals in InGaAs-based NIR camera for (b) cross-polarized polarizer/analyser setup, $P_1=(1,0)^T$ and $P_2=(0,1)$, and (c) parallel-polarized polarizer/analyser setup, $P_1=(0,1)^T$ and $P_2=(0,1)$.  }
   \end{center}
\end{figure}

In order to obtain images of spin-dependent OSH shifts we employed InGaAs-based NIR camera. After two separate measurements of $I_{\sigma^{+}}$ and $I_{\sigma^{-}}$, we calculated $S_{3}=\big(I_{\sigma^{+}}-I_{\sigma^{-}}\big)/\big(I_{\sigma^{+}}+I_{\sigma^{-}}\big) $ from each pixel signals. We examined how OSH shift behaves for $s$-polarization ($y$-polarization) and $p$-polarization ($x$-polarization) of extraordinary refraction beam. In Fig.~\ref{S1} (a) blue and red solid circles correspond to $s$-polarization ($y$-polarization) and $p$-polarization ($x$-polarization), respectively. As shown in Fig.~\ref{S1} (b) and (c), the relative transverse shifts show a sign reversal with the same magnitude, which is different from those observed in air-glass interface. \\

%=======================================================

\end{document}